\documentclass[prc,aps,preprint,showpacs]{revtex4}
\usepackage{graphicx}
\newcommand{\simgeq}{\; \raisebox{-0.4ex}{\tiny$\stackrel
{{\textstyle>}}{\sim}$}\;}

\begin{document}
\title{Interplay between isoscalar and isovector correlations in neutron-rich
nuclei} 

\author{ Ikuko Hamamoto$^{1,2}$ and Hiroyuki Sagawa$^{1,3}$}

\affiliation{
$^{1}$ {\it Riken Nishina Center, Wako, Saitama 351-0198, Japan } \\ 
$^{2}$ {\it Division of Mathematical Physics, Lund Institute of Technology 
at the University of Lund, Lund 22362, Sweden}    \\
$^{3}$ {\it Center for Mathematics and Physics, University of Aizu, 
Aizu-Wakamatsu, Fukushima 965-8580, Japan}  }




\begin{abstract}
The interplay between isoscalar and isovector correlations 
in the 1$^{-}$ states in
neutron-rich (N$\neq$Z) even-even nuclei is studied, taking examples of the
nuclei, $^{22}_{8}$O$_{14}$ and $^{24}_{8}$O$_{16}$.  The excitation modes
explored are isovector dipole and isoscalar compression dipole modes. 
The self-consistent
Hartree-Fock plus the random-phase approximation with the Skyrme interaction,
SLy4, is solved in coordinate space so as to take properly into account the
continuum effect.   The isovector peak induced by isoscalar correlation, the
isoscalar peak induced by isovector correlation, and the possible collective
states made by both isoscalar and isovector correlations, 
(''iS-iV pigmy resonance''), are shown.   
The strong neutron-proton interaction in nuclei can be responsible for
controlling the isospin structure of normal modes.
It is
explicitly shown that in the scattering by isoscalar (isovector) particles on
N$\neq$Z even-even nuclei isovector (isoscalar) strength 
in addition to isoscalar
(isovector) strength may be populated.

\end{abstract}


\maketitle

\newpage

\section{INTRODUCTION} 
The mutual interplay between
isoscalar (IS) and isovector (IV) correlations in neutron-rich nuclei 
is explored in the present work.        
The interplay between IS and IV correlations and the related
collective modes have been 
an attractive and centrally placed 
topic in the study of nuclear structure.     
The IV dipole giant resonance (GR) is the first established giant resonance 
which was found in the photo absorption reaction more than 50 years ago, 
while the corresponding IS partner is the
center of mass motion which is identified as a spurious excitation mode.   
The IS quadrupole correlation is particularly strong in nuclei, and 
IS quadrupole GR is systematically found in experiments, while the 
corresponding IV partner is expected to lie in a higher energy with a broader 
width, but the systematics of the
nature of the IV quadrupole GR is experimentally not yet well established.  
For example, see Ref. \cite{HW01}.     

In the analysis of scattering data by IS particles such as $\alpha$
particles 
it is often assumed that IS particles excite only IS
strength.  This assumption is generally incorrect if the target nuclei have  
N$\neq$Z.  For example, in nuclei with neutron excess IS operators excite 
IS moments, however, the strong neutron-proton forces may tend to keep 
the local ratio of neutrons to protons.  
Then, the presence of neutron excess N$>$Z means 
that IV moments may be excited also by IS particles \cite{BM75}.  
In Fig. 2  of Ref.  \cite{HSZ97} a numerical example of this phenomenon in 
$^{48}$Ca is shown.   Namely, if only IS interaction is taken into account, no
appreciable peak of IV strength is seen around the energy ($\approx$ 16.5 MeV) 
of IS quadrupole giant resonance (QGR).  
In contrast, a sharp IV peak appears at the energy
of the
IS QGR when IV interaction is further included in this nucleus with neutron
excess.  

The phenomena exchanging the above roles of IS and IV excitations 
in the response may be  
expected to be also true.   
Namely, IV operators may produce IS moments generally in nuclei with N$\neq$Z 
except for the case that the IS moment corresponds to the center 
of mass motion.  To study this issue is one of the main points of the present
work. 

Isospin is not an exact but a pretty good quantum-number around the ground state
of nuclei with both N=Z and $|$N$-$Z$|$ $\gg$ 1. 
In N=Z even-even nuclei IS and IV operators excite, roughly
speaking, different
states, since the isospin of the ground state, T$_{0}$, is  
in a good approximation equal to zero.
In contrast, in N$\neq$Z even-even nuclei IS and IV operators
can excite the same states, which have the same isospin 
T$_{0}$=(N$-$Z)/2 as that of of the ground state.   This situation is 
particularly
interesting in nuclei with $|$N$-$Z$|$$\gg$1, 
because the ratio of the population 
of excited states with T=T$_{0}$ to that of higher-lying excited states 
with T=T$_{0}$+1 by IV operator is 
the order of T$_{0}$.  Namely, IV excitations to T=T$_{0}$ states will
consume the major part of the IV strength.

In the present work we choose to study the relation between 
the IV dipole (IVD) 
mode and the IS compression dipole (ISCD) mode.  Both modes have 
spin-parity 1$^{-}$, while various characters besides the isospin (for example, 
the r-dependence of the one-body 
operators or the difference between shape oscillation and compression) 
of the two modes are different.   
In stable nuclei the IVD GR appears energetically  
much lower
than ISCD GR, the major part of which consists of 3$\hbar\omega$ 
excitations.   
In contrast, in neutron drip line nuclei a large fraction of ISCD 
strength is often expected to 
lie much lower than IVD GR \cite{HSZ98}.   In order to study the mutual 
influence by IS and IV modes, it is convenient to study the nuclei, 
in which considerable amounts  
of IVD and ISCD strengths may occur around the same energy region. 
Some such examples are the nuclei, 
$^{22}_{8}$O$_{14}$ and $^{24}_{8}$O$_{16}$, both of which are expected to be
spherical.   This choice of nuclei was stimulated by 
Ref. \cite{AL01} in which the measured photoneutron cross sections for 
$^{22}$O were reported though experimental error bars were large and by Ref. 
\cite{NN17} in which the observation of the bound IS and IV dipole 
excitations in $^{20}_{8}$O$_{12}$ was reported.

In Sec. II the model and the formalism used are briefly summarized, while in
Sec. III numerical results of nuclei, $^{22}$O and $^{24}$O, 
are presented and we
try to study especially the mutual influence between IS and IV correlations in
the N$>$Z nuclei.   In Sec. IV a summary and discussions are given.   

\section{MODEL AND FORMALISM}
We perform the self-consistent Hartree-Fock (HF) 
plus the random-phase approximation
(RPA) with Skyrme interactions, including simultaneously both IS and IV
interactions.   The RPA response function is estimated in coordinate space
so as to take properly into account the continuum effect.   Choosing the Skyrme
interaction SLy4 \cite{EC98}, the model and the formalism are the same as those
employed in Ref. \cite{HSZ98}.   The choice of SLy4 is made 
because, among others, 
the calculated neutron separation energies of $^{22}$O and $^{24}$O are
close to the measured ones and the calculated incompressibility 
in nuclear matter is a realistic value, $K_{0}$= 229.9 MeV.   

Writing the one-body operators   
\begin{equation}
D_{\mu}^{\lambda=1, \,\tau=1} \: = \: \sum_{i}\, \tau_{z}(i) \, r_{i} \, 
Y_{1\mu}(\hat{r}_{i}) 
\label{eq:ivd}
\end{equation}
for IV dipole strength and 
\begin{equation}
D_{\mu}^{\lambda=1, \,\tau=0} \: = \: \sum_{i}\,  
r^{3}_{i} \, 
Y_{1\mu}(\hat{r}_{i}) 
\label{eq:iscd}
\end{equation} 
for IS compression dipole strength, we study the RPA strength function using 
the Green function 
\begin{equation}
S(E) \: \equiv \: \sum_{n} \int \mid \langle n,E_{n} \mid D \mid 0 \rangle 
\mid ^{2} 
\delta (E-E_{n})\, dE_{n}
\: = \: \frac{1}{\pi}\, Im\, Tr (D^{\dagger}\, G_{RPA}(E)\, D) 
\end{equation} 
where $\mid n,E_{n} \rangle$ expresses an excited state 
with energy $E_{n}$ including
both discrete and continuum states.
The transition density for an excited state $\mid \! n \! \rangle$~,
\begin{equation}
\rho_{n0}^{tr}(\vec{r}) \: \equiv \: \langle n \mid \sum_{i=1}^{A} 
\delta(\vec{r}-\vec{r}_i) \mid 0 \rangle \qquad , 
\end{equation}
can be calculated from the RPA response, and the radial transition density
$\rho_{n}^{tr}(r)$ is defined by
\begin{equation}
\rho_{n0}^{tr}(\vec{r}) \: \equiv \: \rho_{n}^{tr}(r) \: Y_{\lambda 
\: \mu}(\hat{r}) \qquad .
\end{equation}

Now, in the case that the numerical calculations of the self-consistent HF+RPA
could be performed with perfect accuracy, 
the spurious center-of-mass state would
be degenerate with the ground state and there would be no spurious component in
the excitation spectra.   However, in numerical calculations it is, in practice,
 difficult to take away completely the dipole strength coming from the spurious
component.   A very small admixture of the spurious component can produce a
pretty large strength especially in the ISCD response.   
Thus, we have to further
eliminate the spurious component in the excitation spectra.   Namely,
as is often done, in the calculation of the IV dipole strength we use the
operator 
\begin{equation}
\bar{D}_{\mu}^{\lambda=1, \,\tau=1} \: = \: -\frac {2N}{A} \, \sum_{i}^{proton} 
\, r_{i} \, 
Y_{1\mu}(\hat{r}_{i}) \; + \; \frac {2Z}{A} \, \sum_{i}^{neutron} \, 
r_{i} \,
Y_{1\mu}(\hat{r}_{i})
\qquad ,
\label{eq:rivd}
\end{equation}
instead of the operator (\ref{eq:ivd}).   In the evaluation of the
strength of IS compression dipole we use the operator
\begin{equation}
\bar{D}_{\mu}^{\lambda=1, \,\tau=0} \: = \: \sum_{i}^{A} 
\, (r_{i}^3 \, - \, \eta \, r_{i}) \,  
Y_{1\mu}(\hat{r}_{i}) 
\label{eq:riscd}
\end{equation}
where ~$\eta= \frac {5}{3} <\! r^2 \!>$~, instead of the operator
(\ref{eq:iscd}).   

\section{NUMERICAL RESULTS}
Oxygen isotopes are generally expected to be spherical, except for very
neutron-deficient 
isotopes such as $^{12}_{8}$O$_{4}$, which is the mirror nucleus of
$^{12}$Be.  Here we study $^{22}$O and $^{24}$O, both of which are treated as
spherical j-j shell closed nuclei.   

First, we check whether in our present model we obtain the experimentally
reported amount of ISCD strength in a bound state of $^{16}_{8}$O$_{8}$.   
From the analysis of ($\alpha, \alpha'$) scattering it is 
reported that the 1$^-$ and T=0 state at 7.12 MeV consumes about 4 percent of
the ISCD energy-weighted sum-rule (EWSR) \cite{HD81}, while about twice 
the strength on the same state was
reported from the analysis of the electron scattering data 
\cite{TJD73}.   Applying our model to $^{16}$O, we obtain the ISCD strength of
about 2 percent of ISCD EWSR as a sum of the strength of 
two bound RPA solutions at 10.4 and
10.8 MeV.   Therefore, we expect that our calculation of ISCD spectrum can be
reasonably realistic.   

\subsection{The nucleus $^{22}$O}
Using the SLy4 interaction, we obtain the neutron and proton separation
energies,  $S_n$ = 7.25 MeV and $S_p$ = 22.29 MeV,
compared with the experimental values, $S_n$ = 6.85 MeV and $S_p$ = 23.26 MeV. 

In Fig. 1 we show the calculated RPA strength for ISCD operator.  The
solid curve is the strength when both IS and IV interactions are included 
in RPA,
while the strength obtained by including only IS interaction in RPA is denoted
by the dotted curve.  No RPA solution is found below the threshold energy. We
note the following points.  
(a) As was already found in Ref. \cite{HSZ98} concerning the ISCD response in
neutron drip line nuclei, we very often 
obtain a large portion of ISCD strength in the
energy interval of several MeV above the threshold.  
This large strength appearing
in the energy much lower than the energy of the ISCD GR, which is recognized 
as a very broad ''resonance''
found for Ex$\simgeq$24 MeV in Fig. 1, comes from the possible presence of
occupied weakly-bound low-$\ell$ neutron orbits together with the strong 
$r$-dependence ($r^{3}$) of the ISCD operator.  This point is later discussed in
more detail in subsec. III. B related to Figs. 5 and 6 for $^{24}$O.   
(b) When IV interaction is included on the top of IS interaction, the
heights of many lower-lying IS peaks become lower and the peak energies  
may shift to slightly higher energies via the IV components contained 
in those IS peaks, due to the repulsive nature of the IV interaction.   
(c) There are some peaks denoted by the solid curve, which may not be understood
in the way of the above (b).  An example is 
the broad peak around 18.5 MeV in the
solid curve. The nature of the IS peaks around 14.0 and 18.5 MeV, which have no
trivial corresponding peaks in the dotted curve, is later discussed in
comparison of Fig. 1 with Fig. 2.   

In Fig. 2 the calculated RPA strength for IVD operator is shown as a
function of excitation energy.  The strength, in which both IS and IV
interactions are included in RPA, is denoted by the solid curve, while the
strength obtained by including only IV interaction in RPA is expressed by the
long-dashed curve.  No RPA solution is found below the threshold energy.  
The unperturbed particle-hole (p-h) strength, in which particle levels are in
the continuum, is shown by the dotted curve, while the energies of seven
unperturbed p-h excitations, in which particle levels are bound states, are
shown by vertical arrows.  There are two such neutron excitations and five such 
proton excitations.  Since the contributions to the unperturbed strength by the
excitations from such bound states to bound states have a dimension different 
from the dimension of the contributions by the excitations 
from bound states to continuum states 
due to the unavoidable difference between the
normalization of bound states and that of continuum states, the two kinds of
contributions cannot be shown in the same figure in a simple manner.   So, we
 just show the p-h energies of the excitations from bound states to bound
states.   It is noted that the RPA result always contains the contributions 
by both 
excitations from bound to bound states and those from bound to continuum
states.

It is known that in light nuclei such as oxygen isotopes even IVD GR does not
appear as a clean large resonance, instead, it splits into several
considerable peaks.  Therefore, roughly speaking, the peaks in the region of
Ex$>$18 MeV in Fig. 2 may be regarded as IVD GR of $^{22}$O.   
It is seen that the
main peaks belonging to the IVD GR hardly change 
when IS interaction is included on
the top of IV interaction.   That means, those peaks hardly have IS components
even in this nucleus with neutron excess.   In contrast, several IV peaks
between the threshold energy and about 16 MeV, which
are obtained by including only IV interaction, seem to be reorganized to
approximately one relatively broad peak 
with the peak energy around 14 MeV, when IS
interaction is included.  The reorganization is certainly made through the IS 
components contained in those several IV peaks.

In Fig. 3 the radial transition densities multiplied by $r^{2} r^{3}$ for IS
and $r^{2} r$ for IV, which are calculated at Ex = 18.5 MeV 
in the continuum, are shown in
arbitrary units, as a function of radial variable.  Here the multiplied factor
$r^2$ expresses the volume element which appears in the integration over 
$r$.   The total IV (= neutron $-$ proton) transition density denoted by the
thin solid curve shows, roughly speaking, a typical shape of the IVD GR
transition density with a peak around the nuclear surface.  An interesting
feature found in Fig.3 is that the $r$-value of the (negative) inner peak
(around 3.4 fm) of the total ISCD transition density 
is almost the same as that of the peak of the total IV transition density.  
This fact together with the creation of the relatively broad
peak of ISCD strength when IV interaction is included on the top of IS
interaction leads to the interpretation that the relatively broad ISCD peak 
around 18.5 MeV in Fig. 1 is
induced by the IV peak at the same energy via the IV component 
contained in the ISCD peak.      

In Fig. 4 the radial transition densities multiplied by $r^{2} r^{3}$ for IS and
$r^{2} r$ for IV calculated at Ex = 14.0 MeV in the continuum are shown in
arbitrary units, as a function of radial variable.   
The total IV transition density has a peak around 4 fm
clearly outside the nuclear surface, at which r-value the total IS (= neutron 
$+$ proton) transition density expressed by the thick solid curve has
approximately a node in contrast to the case of Fig. 3.   Then, we may call the
relatively broad peak around Ex = 14 MeV, which appears in the solid curve 
expressing the
final RPA result in both Fig. 1 and Fig. 2, pigmy resonance with both isoScalar
and isoVector correlations (''iS-iV pigmy resonance'').   
The pigmy resonance is interpreted as neither the IS
strength induced by a strong IV peak nor the IV strength induced 
by a strong IS peak, due
to the presence of neutron excess.   It is a relatively broad resonance, 
of which the
energy is much lower than the energies of 
both IVD GR and ISCD GR, but it gathers the
collectivity of low-lying IS and IV strengths.  
The IV dipole strength in the region of Ex = (11-16.5) MeV consumes 11.4
percent of the energy-weighted sum-rule (EWSR) and 16.6 percent of the
non-energy-weighted sum-rule (NEWSR).  For ISCD mode (see Fig. 1) we find even 
a larger portion of the contribution to the sum rule from the same energy 
region:
18.0 percent for EWSR and 25.9 percent for NEWSR.  
Such an iS-iV pigmy 
resonance will be found neither in stable nuclei because the peaks with
appreciable strength of ISCD appear in the energy region different from that of
IVD 
nor in N=Z nuclei because the isospin of ISCD states is 0 while that of IVD 
states is 1.   

In the continuum RPA, we have no explicit values for the amplitudes of
respective p-h components.  Nevertheless, the collective nature of a given
resonance can be seen from the contribution to the sum rule and the number
of available configurations.
For IV 1$^-$ states in Fig. 2 we show the unperturbed response of the
excitation to the continuum (by the dotted curve) and the p-h energies of
the excitation to bound states (by arrows).   In the energy region of
Ex = 10-18 MeV there are two resonance states and six bound p-h states. 
Starting from the lowest-energy peak, they are assigned as
 (a) neutron (2s$_{1/2}$, 1p$_{1/2}^{-1}$), 
 (b) neutron (1f$_{7/2}$, 1d$_{5/2}^{-1}$) at 12.2 MeV with the width 1.22 MeV, 
 (c) proton (2s$_{1/2}$, 1p$_{1/2}^{-1}$), 
 (d) proton (1d$_{5/2}$, 1p$_{3/2}^{-1}$), 
 (e) proton (1d$_{3/2}$, 1p$_{1/2}^{-1}$), 
 (f) neutron (1d$_{3/2}$, 1p$_{1/2}^{-1}$) at 16.8 MeV with the width 83 keV, 
 (g) neutron (2s$_{1/2}$, 1p$_{3/2}^{-1}$), and  
 (h) proton (2s$_{1/2}$, 1p$_{3/2}^{-1}$).   
They get some finite width as shown by the long-dashed curve between
11 and 18 MeV due to the coupling to the continuum in RPA with
IV interaction.  The main configurations of respective peaks expressed by the
long-dashed curve can be easily
assigned to respective unperturbed peaks, because corresponding
unperturbed peaks were pushed only slightly to higher energy due to the
repulsive nature of IV interaction.  Some strength of those low-lying
unperturbed peaks must be shifted to the high-lying giant resonance due to
the repulsive IV interaction.  Nevertheless, the correspondence between
the unperturbed and perturbed (only by IV interaction) peaks 
in the energy region of 11-17 MeV can
be clearly seen in Fig. 2.
When we further introduce IS interaction on the top of IV interaction, we
obtain iS-iV pigmy resonance, of which the main component comes from the
configurations above (in particular, five lower-energy configurations) 
and collect
most of the strength around Ex=14 MeV.  As is seen in Fig.4, there are
substantial contributions also from proton p-h excitations.

It is further noted that, for example, in Fig. 4 the $r$-values (about 3.3 fm
and 9.2 fm), at which the curve of  
''neutron IS $r^{3}$'' becomes zero, are clearly different from those, 
at which the curve of ''neutron IV $r$'' is zero.  
Since the multiplied factors $r^{3}$ and $r$ do not make such nodes, 
these nodes come
from the nodes of the transition densities.  Namely, the relevant calculated
radial transition densities of ''neutron IS $r^3$'' and ''neutron IV $r$'' can
be different in a subtle manner, though both neutron transition densities are
calculated at the same energy.   
Solving the continuum RPA makes it possible to obtain the possible 
difference.          
If the RPA calculation were performed, for example, 
by expanding the wave functions in terms of a finite number of 
harmonic-oscillator bases, the
difference  
would not have appeared, because the shapes (then, the nodes) of 
the radial transition 
densities for protons and neutrons  
at a given energy are uniquely determined since the bases
have no intrinsic widths.
In contrast, if two peaks, of which one is IS while the other is IV, 
with finite widths occur energetically 
close to each other, the transition density 
at a given energy between the two peaks 
can be a mixture of the components of the two peaks.  
If so, the neutron (and proton) transition 
density for IS operator can be different from that for IV operator.

\subsection{The nucleus $^{24}$O}
Using the SLy4 interaction, we obtain $S_n$ = 4.88 MeV and $S_p$ = 24.62 MeV,
compared with the experimental values $S_n$ = 4.19 MeV and $S_p$ = 27.11 MeV.  

By the solid curve in Fig. 5 we show the RPA strength, 
in which both IS
and IV interactions are included in RPA, while the RPA
strength with only IS interaction is shown by the dotted curve.    
No RPA solution is found below the threshold energy.   
The ISCD GR is recognized as
a very broad bump with a peak at slightly above 30 MeV.  As in the case of
$^{22}$O, in the energy region much lower than ISCD GR we obtain many strong
peaks.  In particular, we note a huge calculated peak at 5.8 MeV with a width
of a few MeV.  The main component of the 5.8 MeV peak is the neutron (p-h) =
($2p_{3/2} 2s_{1/2}^{-1}$) configuration.  

In Fig. 6 
the radial transition density
multiplied by $r^{2} r$, which corresponds to the center of mass motion, is
plotted by the solid curve.  The integration of the solid curve from $r$=0 to
infinity is zero if the 5.8 MeV $1^-$ state contains no spurious component due
to the center of mass motion.  
In other words, only after the integration over $r$ the degree of the
contamination of the spurious component can be seen.  Radial transition
densities of neutrons and protons are determined by the properties of 
excitation
modes at given energies and are the same, for example, in the spurious strength
(corresponding to the center of mass motion) and the ISCD strength.  The
difference between the contributions to the two kinds of strengths comes only
from the different multiplication factors depending on $r$.
It is noted that  
the solid curve in Fig. 6 extends considerably to the region far outside 
the size of the nucleus.  If we multiply the
solid curve further by $r^2$ so as to obtain the ISCD strength, 
it is seen that the main contribution to the
ISCD strength comes from the region far outside the nucleus.  This
peculiar situation comes from the characteristic behavior of the main component,
the neutron (p-h) = ($2p_{3/2} 2s_{1/2}^{-1}$) configuration.  
The neutron $2s_{1/2}$ level,
which is the last occupied orbit, has an energy of $-$4.88 MeV in the HF
calculation, while the 
$2p_{3/2}$ level is slightly before becoming a one-particle resonant level.  
Consequently, the contribution to the radial transition density by the p-h
configuration has an extended tail
to the outside of the nucleus.  Then, since the ISCD operator has the strong
$r$-dependence ($r^{3}$), this main component produces a very large ISCD 
strength making a large contribution to the ISCD sum-rule.  
Thus, it can be said that the strong peak at 5.8 MeV is not really 
the result of
large collectivity but the peak is strong due to the unique behavior of 
the neutron low-$\ell$ wave functions.   

In Fig. 7 the calculated RPA strength with the full interactions 
for the IVD operator is shown by the
solid curve, while the RPA strength obtained by including only IV
interaction in RPA is shown by the long-dashed curve.  The unperturbed p-h
strength for the IVD operator, in which particle states are in the continuum, 
is denoted by the dotted curve expressed as ''unperturbed''.  
The energies of the unperturbed
p-h excitations, in which particle levels are bound states, are shown by
vertical arrows.  There are five such proton p-h excitations within the energy
range of Fig. 7 and two such neutron
p-h excitations.   All unperturbed p-h  excitations, in which particle states
are in the continuum as well as bound states, contribute to the RPA solutions.  
The major component of the small bump in the solid curve, of which the peak is
around 6.5 MeV, is the neutron (p-h) = ($2p_{3/2} 2s_{1/2}^{-1}$) configuration.
  Since the
IVD operator has the mild $r$-dependence ($r$), the response is not so strong as
the ISCD response shown in Fig. 5, in spite of the fact that 
the IVD strength comes mainly from the outside
of the nuclear surface.  

The relation between the IV and IS correlations, which was obtained in
subsec. III. A for $^{22}$O, is found also in $^{24}$O in a very similar manner.
For example, the relatively sharp ISCD peak around 16.7 MeV in Fig. 5 can be
interpreted as the one 
induced by the IVD peak at the same energy, as in the case of the 18.5 MeV ISCD
peak of $^{22}$O in Fig. 1.   Similarly, several IVD peaks between 11 and 16
MeV, which are obtained by including only IV interaction in RPA, 
are reorganized into, roughly speaking, 
one relatively broad peak with a center around 14 MeV, 
when IS interaction is further
included in RPA.  
This is what we have called ''iS-iV pigmy resonance'' in the subsec. III. A.   
We have found that the structure of the pigmy resonance in $^{24}$O is very
similar to that in $^{22}$O.   
    
\section{SUMMARY AND DISCUSSIONS}
The IV peak induced by IS correlation, the IS peak induced by IV
correlation, and the possible collective states made by both IS and IV
correlations in N$\neq$Z nuclei are studied  
by taking $^{22}$O and $^{24}$O.  
The spin-parity of the states is 1$^{-}$, consequently, 
the excitation mode in the
IV channel is IVD which has been well studied both experimentally and
theoretically, while that in the IS channel is ISCD which has been theoretically
pretty well studied though the related experimental information is still rather
limited.   Those light neutron-drip-line nuclei are chosen to study the present
topic, 
mainly because the
peaks with an appreciable amount of the strength of ISCD and those of IVD may be
expected in the same energy region.  

Taking the example of QGR in $^{48}$Ca, 
in Ref. \cite{HSZ97} it is shown that a
sharp appreciable IV peak appears at the energy of the IS QGR 
when both IS and IV
interactions are taken into account while no peak of such IV strength is
found if only IS interaction is included.   This is one of the typical
phenomena, in which IS collective modes in nuclei with N $>$ Z 
may contain some IV
strength.  

As examples of the phenomena in which IV collective modes induce 
some IS strength
in nuclei with N  $>$ Z, in the present work we have shown the peak at 18.5 MeV
in $^{22}$O as well as the peak at 16.7 MeV in $^{24}$O. 

Furthermore, in both $^{22}$O and $^{24}$O we have obtained the relatively-broad
medium-size peak around 14 MeV, which is much lower than both ISCD GR and
IVD GR and several
MeV higher than the threshold energy.  
The peak contains both IS and IV correlations.  
We call the peak ''iS-iV pigmy resonance''.   Indeed,
the IV component of the pigmy resonance is clearly the result of reorganizing
several IV peaks that are obtained by including only IV interaction in RPA. 
The reorganization could occur via the IS component contained in those several 
IV peaks.   

In the present work together with the  IS QGR example of $^{48}$Ca 
in Ref. \cite{HSZ97} 
it is explicitly shown that in the scattering 
by IS (IV) particles some IV (IS) strength in addition to IS (IV) strength 
may be 
populated, in the case that target nuclei have N$\neq$Z.    
This can be understood as the result of the strong neutron-proton force, by
which the isospin structure of normal modes is controlled.    
In a given nucleus with N$\neq$Z the ratio of the population of
IV strength to that of IS strength by IS (or IV) particles seems to depend very
much on the structure of individual states.

As seen in Figs. 1 and 5, one may expect a large amount of ISCD strength in the
energy region above the threshold energy and much lower than ISCD GR, when
weakly-bound low-$\ell$ neutron levels are occupied in the ground state of
nuclei and the particle state of the p-h components 
is a resonance (or almost
resonance) in the continuum.   
If the particle state is bound, one may say that 
it is due to the so-called halo phenomenon 
in both
particle and hole states.   It may be very interesting to observe this
phenomenon experimentally, though the actual strength of this kind of peak will
depend rather sensitively on the actual one-particle energies of the particle
and hole levels.
 
By using the material in the present paper, we have shown an important strong
point of the method of solving the continuum RPA.   
Namely, when two (IS and IV) peaks
occur in the continuum within respective widths, the transition
densities of neutrons and protons for a given energy within the overlapped
widths may not be uniquely
determined, instead, they may depend on the (IS or IV) 
field operated externally.  
In contrast, if one uses the expansion of wave functions in terms
of a finite number of basis wave-functions which have no intrinsic width, 
one can obtain only the
transition densities of neutrons and protons 
which are determined uniquely for a given energy.     
 
H. S. appreciates the partial support by JSPS KAKENHI Grant Numbers 16H02179.

\newpage

\newpage
\noindent
{\bf\large Figures}\\
\begin{figure}[htp]
\vspace{2cm}
\includegraphics[scale=0.45,width=8cm,angle=-90]{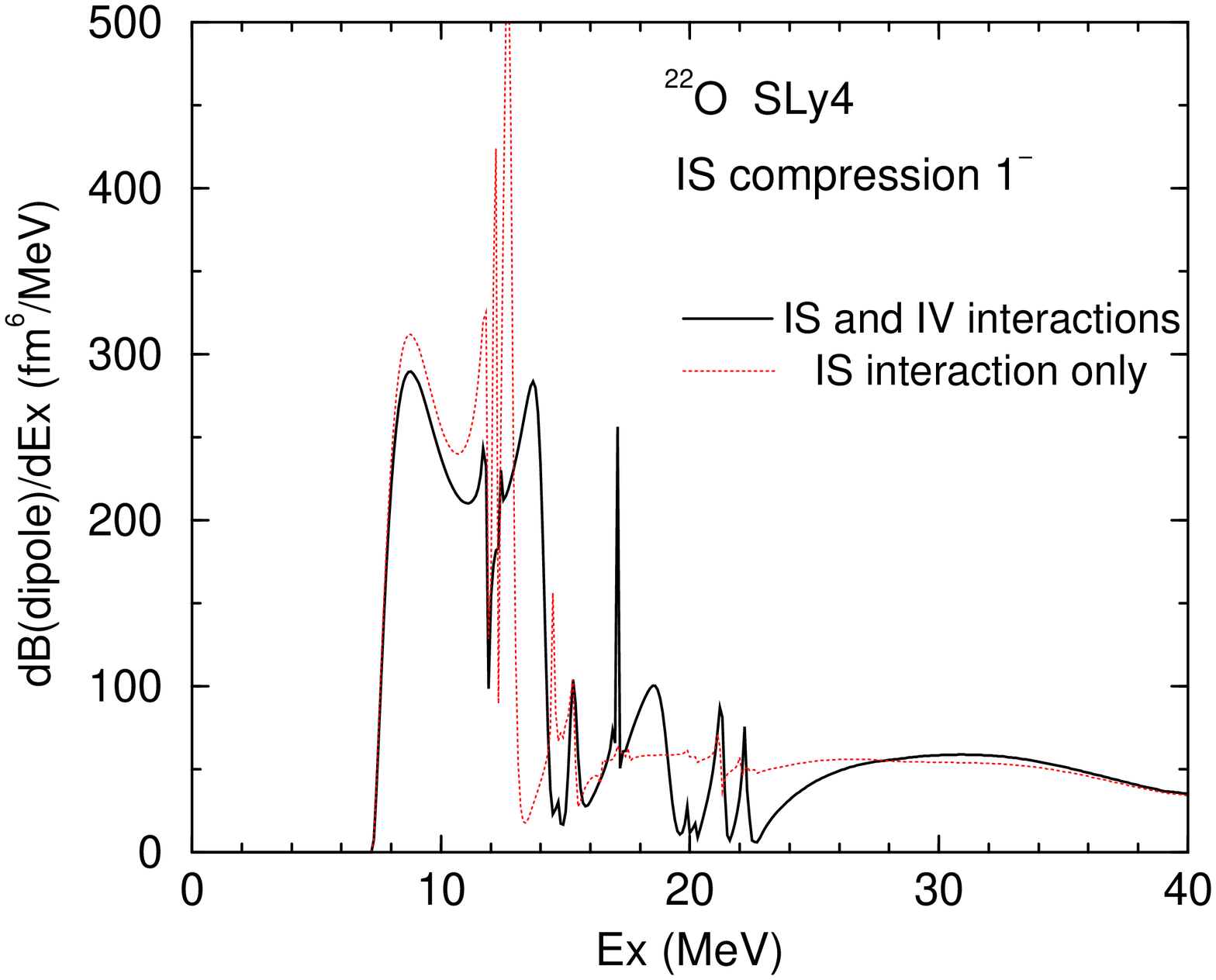}
\caption{(Color online)The RPA strength of isoscalar compression dipole (ISCD)  
of $^{22}$O calculated by using the SLy4 interaction as a function of 
excitation energy (Ex).
The solid curve expresses the strength, in which both IS and IV interactions are
included in RPA, while the dotted curve is obtained by including only IS
interaction.  No RPA solution is obtained below the threshold energy.
\label{fig1}}
\end{figure}

\begin{figure}[htp]
\vspace{2cm}
\includegraphics[scale=0.45,width=8cm,angle=-90]{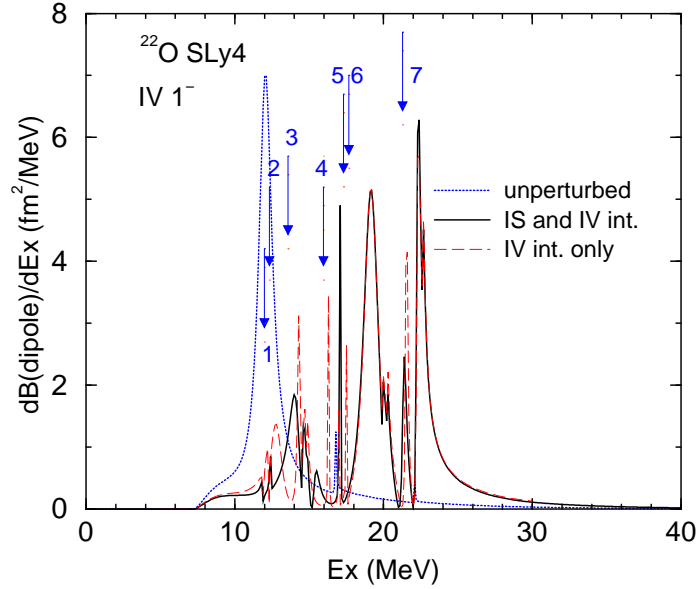}
\caption{(Color online)The RPA strength of isovector dipole (IVD) of $^{22}$O
calculated by using the SLy4 interaction as a function of excitation energy. 
The solid curve expresses the strength obtained by including both IS and IV
interactions in RPA, while the long-dashed curve is obtained by including only
IV interaction in RPA. No RPA solution is obtained below the threshold
energy.
The unperturbed p-h strength, of which particle
states are in the continuum, is denoted by the dotted curve, while the energies
of the unperturbed p-h excitations, of which particle states are bound states,
are denoted by vertical arrows.  
The numbered arrows correspond to the following (p-h) configurations: 
$(1) (2s_{1/2}1p_{1/2}^{-1})_{\nu}, (2) (2s_{1/2}1p_{1/2}^{-1})_{\pi}, (3)
   (1d_{5/2}1p_{3/2}^{-1})_{\pi}, (4) (1d_{3/2} 1p_{1/2}^{-1})_{\pi}, 
   (5) (2s_{1/2} 1p_{3/2}^{-1})_{\nu},$ 
   $(6) (2s_{1/2}1p_{3/2}^{-1})_{\pi}, $
    and $(7)  
  (1d_{3/2}1p_{3/2}^{-1})_{\pi}$, respectively.  
The large unperturbed strength with a peak around 12 MeV denoted by the dotted
curve is the neutron (p-h) = ($1f_{7/2}1d_{5/2}^{-1}$) resonance.
\label{fig2}}
\end{figure}

\begin{figure}[htp]
\vspace{2cm}
\includegraphics[scale=0.35,width=8cm,angle=-90]{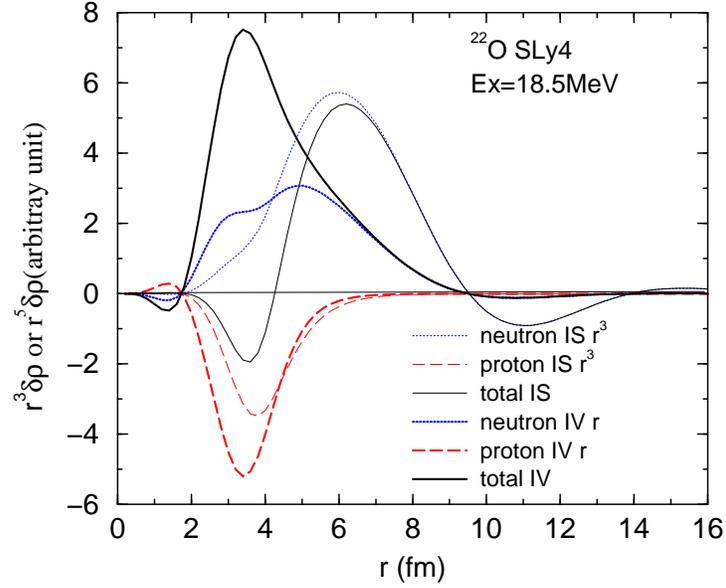}
\caption{(Color online)Radial transition densities multiplied by $r^{2} r^{3}$ for ISCD and those 
multiplied by $r^{2} r$ for IVD, which are calculated at Ex
= 18.5 MeV.  Both IS and IV interactions are included.  
Note the relatively broad peak around 18.5 MeV denoted by the solid curve 
in Fig. 1 and the relatively broad peak with the center
around 19 MeV expressed by the solid curve in Fig. 2.  Thick curves are for
ISCD, while thin curves are for IVD.  Dotted curves are for neutrons, while
long-dashed curves are for protons.  
\label{fig3}}
\end{figure}

\begin{figure}[htp]
\vspace{2cm}
\includegraphics[scale=0.35,width=8cm,angle=-90]{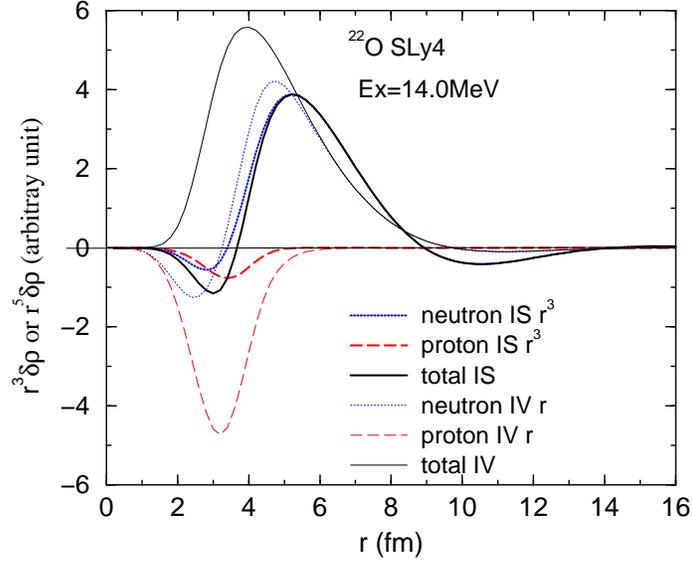}
\caption{(Color online)Radial transition densities multiplied by $r^{2} r^{3}$ for ISCD and those 
multiplied by $r^{2} r$ for IVD, which are calculated at Ex
= 14.0 MeV.  Both IS and IV interactions are included.  
Note that the relatively 
broad peaks
expressed by the solid curves appear around 14 MeV in both Figs. 1 and 2. Thick
curves are for ISCD, while thin curves are for IVD.   Dotted curves are for
neutrons, while long-dashed curves are for protons.        
\label{fig4}}
\end{figure}

\begin{figure}[htp]
\vspace{2cm}
\includegraphics[scale=0.35,width=8cm,angle=-90]{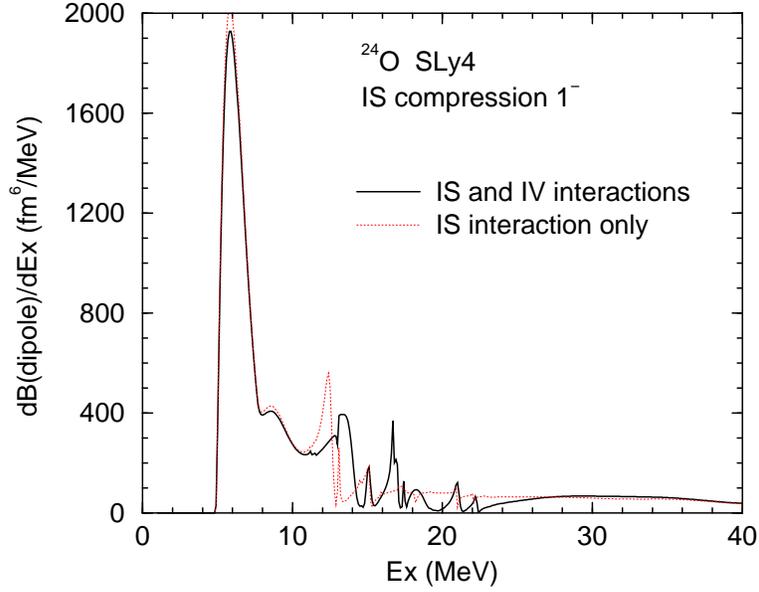}
\caption{(Color online)The RPA strength of isoscalar compression dipole of $^{24}$O
calculated by using the SLy4 interaction as a function of excitation energy. 
The solid curve expresses the strength, in which both IS and IV interactions are
included in RPA.  The strength, in which only IS interaction is included in
RPA, is denoted by the dotted curve.  No RPA solution is obtained below the
threshold energy.  
\label{fig5}}
\end{figure}

\begin{figure}[htp]
\vspace{2cm}
\includegraphics[scale=0.35,width=8cm,angle=-90]{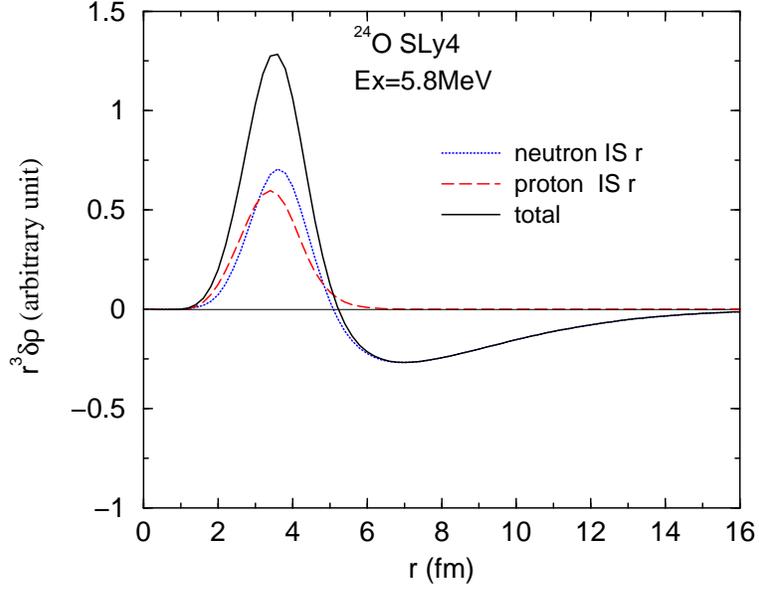}
\caption{(Color online)Radial transition densities that are multiplied by $r^{2} r$ and expressed 
in arbitrary unit, which
are calculated at Ex = 5.8 MeV of $^{24}$O
as a function of radial variable.  Note that the
peak energy of the huge response just above the threshold in Fig. 5 is 5.8 MeV. 
The neutron (proton) radial transition density is expressed by the dotted
(long-dashed) curve, and the sum of the dotted and long-dashed curves is 
denoted by the solid curve.  
\label{fig6}}
\end{figure}

\begin{figure}[htp]
\vspace{2cm}
\includegraphics[scale=0.35,width=8cm,angle=-90]{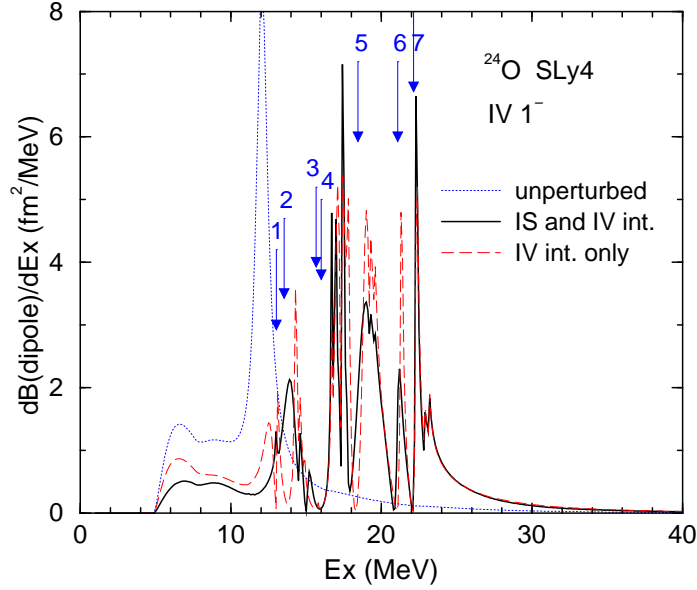}
\caption{(Color online)
The RPA strength of isovector dipole of $^{24}$O calculated by
using the SLy4 interaction as a function of excitation energy.  The strength, in
which both IS and IV interactions are included in RPA, is shown by the solid
curve, while the strength, in which only IV interaction is included in RPA,
is denoted by the long-dashed curve.  No RPA solution is obtained below the
threshold energy.  The dotted curve expresses the unperturbed strength, in which
particle states in p-h excitations are in the continuum.  The energies of the
unperturbed p-h excitations, of which particle states are bound, are denoted by
vertical arrows.   
The numbered arrows correspond to the following (p-h) configurations: 
(1)
 $(2s_{1/2}1p_{1/2}^{-1})_{\pi}, (2)  (1d_{5/2}1p_{3/2}^{-1})_{\pi}, 
 (3) (1d_{3/2}1p_{1/2}^{-1})_{\pi},  (4) (1d_{3/2}1p_{1/2}^{-1})_{\nu}, 
(5) (2s_{1/2}1p_{3/2}^{-1})_{\pi},$  $(6) (1d_{3/2}1p_{3/2}^{-1})_{\pi},$ and   
$(7) (1d_{3/2}1p_{3/2}^{-1})_{\nu},$ respectively.
\label{fig7}}
\end{figure}
\end{document}